\documentclass[a4paper]{jpconf}
\usepackage{graphicx}
\newcommand{\lesssim}{\raisebox{0.3mm}{\em $\, <$} \hspace{-3.2mm}
\raisebox{-1.4mm}{\em $\sim \,$}}
\begin{document}
\title{Constraints on the flavor-dependent non-standard interaction in propagation from atmospheric neutrinos}

\author{Osamu Yasuda}

\address{Department of Physics, Tokyo Metropolitan University,
Minami-Osawa, Hachioji, Tokyo 192-0397, Japan}

\begin{abstract}
The sensitivity of the atmospheric neutrino
experiments to the non-standard flavor-dependent
interaction
in neutrino propagation is studied under the
assumption that the only nonvanishing components
of the non-standard matter effect are
the electron
and tau neutrino components $\epsilon_{ee}$,
$\epsilon_{e\tau}$, $\epsilon_{\tau\tau}$
and that the tau-tau component satisfies
the constraint
$\epsilon_{\tau\tau}=|\epsilon_{e\tau}|^2/(1+\epsilon_{ee})$
which is suggested from the high energy
behavior for atmospheric neutrino data.
\end{abstract}

\section{Introduction}

Neutrino oscillations in the standard three-flavor scheme
are described by three mixing angles,
$\theta_{12}$, $\theta_{13}$, $\theta_{23}$, one CP phase $\delta$,
and two independent mass-squared differences, $\Delta m^2_{21}$ and
$\Delta m^2_{31}$.
Thanks to the recent progress of the experiments with solar, atmospheric,
reactor and accelerator neutrinos,
the three mixing angles and the
two mass-squared differences
have been determined.
The only oscillation parameters which are
still undetermined are
the value of the CP phase $\delta$ and the sign of
$\Delta m^2_{31}$ (the mass hierarchy).
In the future neutrino long-baseline experiments
with intense neutrino beams 
the sign of $\Delta m^2_{31}$ and
$\delta$ are expected to be determined\,\cite{Abe:2014oxa,Adams:2013qkq}.
As in the case of B factories,
such high precision measurements will enable us to
search for deviation from the standard three-flavor oscillations
(see, e.g., Ref.\,\cite{Bandyopadhyay:2007kx}).
Among such possibilities,
in this talk, we will discuss
the effective non-standard neutral current
flavor-dependent neutrino interaction with
matter,
given by
\begin{eqnarray}
{\cal L}_{\mbox{\rm\scriptsize eff}}^{\mbox{\tiny{\rm NSI}}} 
=-2\sqrt{2}\, \epsilon_{\alpha\beta}^{fP} G_F
(\overline{\nu}_\alpha \gamma_\mu P_L \nu_\beta)\,
(\overline{f} \gamma^\mu P f'),
\label{NSIop}
\end{eqnarray}
where $f$ and $f'$ stand for fermions (the only relevant
ones are electrons, u and d quarks),
$G_F$ is the Fermi coupling constant, and $P$ stands for
a projection operator that is either
$P_L\equiv (1-\gamma_5)/2$ or $P_R\equiv (1+\gamma_5)/2$.
If the interaction (\ref{NSIop}) exists, then
the standard matter effect
is modified.
We will discuss atmospheric neutrinos
which go through the Earth, so we make an approximation
that the number densities of electrons ($N_e$),
protons, and neutrons are equal.
Defining
$\epsilon_{\alpha\beta}
\equiv \sum_{P}
\left(
\epsilon_{\alpha\beta}^{eP}
+ 3 \epsilon_{\alpha\beta}^{uP}
+ 3 \epsilon_{\alpha\beta}^{dP}
\right)$,
the hermitian $3 \times3 $ matrix of the matter potential becomes
\begin{eqnarray}
{\cal A}\equiv A\left(
\begin{array}{ccc}
1+ \epsilon_{ee} & \epsilon_{e\mu} & \epsilon_{e\tau}\\
\epsilon_{\mu e} & \epsilon_{\mu\mu} & \epsilon_{\mu\tau}\\
\epsilon_{\tau e} & \epsilon_{\tau\mu} & \epsilon_{\tau\tau}
\end{array}
\right),
\label{matter-np}
\end{eqnarray}
where $A\equiv\sqrt{2}G_FN_e$ stands for the matter effect
due to the charged current interaction in the standard case.
With this matter potential, the Dirac equation for
neutrinos in matter becomes
\begin{eqnarray}
i {d \over dx} \left( \begin{array}{c} \nu_e(x) \\ \nu_{\mu}(x) \\ 
\nu_{\tau}(x)
\end{array} \right) = 
\left[  U {\rm diag} \left(0, \Delta E_{21}, \Delta E_{31}
\right)  U^{-1}
+{\cal A}\right]
\left( \begin{array}{c}
\nu_e(x) \\ \nu_{\mu}(x) \\ \nu_{\tau}(x)
\end{array} \right),
\label{eqn:sch}
\end{eqnarray}
where $U$ is the leptonic mixing matrix,
and $\Delta E_{jk}\equiv\Delta m_{jk}^2/2E\equiv (m_j^2-m_k^2)/2E$.

Constraints on $\epsilon_{\alpha\beta}$
have been discussed by many authors
(see, e.g., Refs.\,\cite{Bandyopadhyay:2007kx},
\cite{Fukasawa:2015jaa} and references therein),
and it is known that the bounds on 
$\epsilon_{ee}$, $\epsilon_{e\tau}$ and $\epsilon_{\tau\tau}$ are
much weaker than those on $\epsilon_{\alpha\mu}~(\alpha=e,\mu,\tau)$.
Taking into account the various constraints,
in the present talk we take the ansatz
\begin{eqnarray}
{\cal A}= A\left(
\begin{array}{ccc}
1+ \epsilon_{ee}~~ & 0 & \epsilon_{e\tau}\\
0 & 0 & 0\\
\epsilon_{e\tau}^\ast & 0 & ~~|\epsilon_{e\tau}|^2/(1 + \epsilon_{ee})
\end{array}
\right),
\label{ansatz}
\end{eqnarray}
and analyze the sensitivity to
the parameters $\epsilon_{\alpha\beta}~(\alpha,\beta=e,\tau)$
of the atmospheric neutrino experiment at
Superkamiokande (SK) and the future Hyperkamiokande (HK)
facility\,\cite{Abe:2011ts}.
This talk is based on the work\,\cite{Fukasawa:2015jaa}
and the readers are referred to Ref.\,\cite{Fukasawa:2015jaa}
for details.

\section{The constraint of the Superkamiokande atmospheric
neutrino experiment on $\epsilon_{ee}$ and $|\epsilon_{e\tau}|$
\label{atmospheric-sk}}

First let us discuss the constraint of the
SK atmospheric neutrino experiment on
the non-standard interaction in propagation with the ansatz
(\ref{ansatz}).
The independent degrees of freedom
in addition to those in the standard
oscillation scenario are
$\epsilon_{ee}$, $|\epsilon_{e\tau}|$
and $\mbox{\rm arg}(\epsilon_{e\tau})$.
We have performed a $\chi^2$ analysis
for the SK atmospheric neutrino data 
for 4438 days\,\cite{Abe:2014gda}.

\begin{figure}
\vspace*{-5mm}
\includegraphics[scale=0.4]{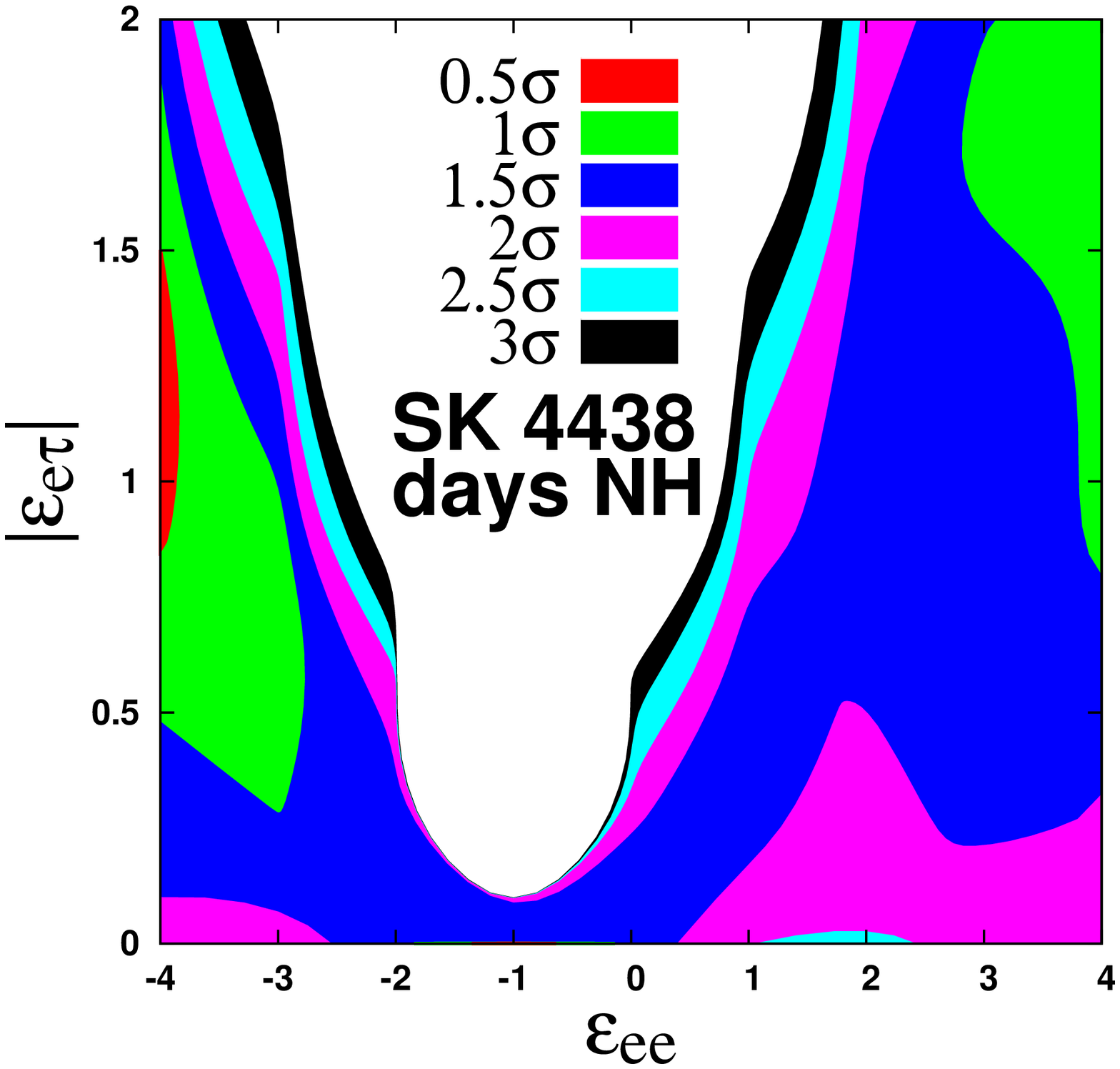}
\includegraphics[scale=0.4]{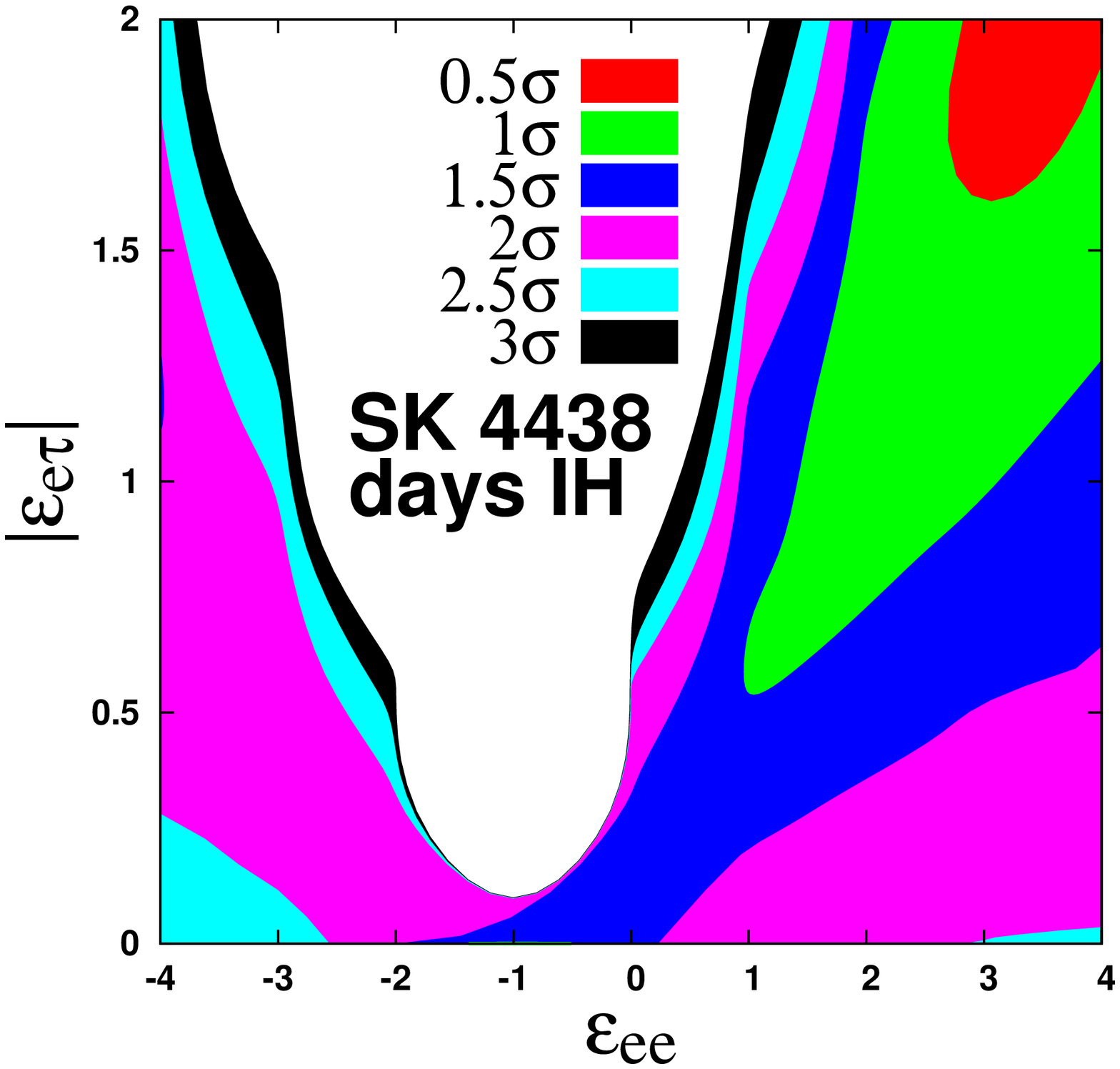}
\vspace*{5mm}
\caption{
The allowed regions
in the ($\epsilon_{ee}$, $|\epsilon_{e\tau}|$) plane
from the SK atmospheric neutrino data for a
normal (inverted) mass hierarchy.
}
\label{fig:fig1}
\end{figure}

The result for the Superkamiokande data
for 4438 days is given in Fig.\,\ref{fig:fig1},
where the allowed region is plotted
in the ($\epsilon_{ee}$, $|\epsilon_{e\tau}|$) plane.
It is understood that, at each point in
this plane, $\chi^2$ is marginalized with respect to
the parameters $\theta_{23}$, $|\Delta m^2_{31}|$,
$\delta$ and $\mbox{\rm arg}(\epsilon_{e\tau})$.
All other oscillation parameters are fixed
for simplicity and the reference values used here are
the following:
$\sin^22\theta_{12}=0.86$,
$\sin^22\theta_{13}=0.1$,
$\Delta m^2_{21}=7.6\times 10^{-5}\mbox{\rm eV}^2$.
The best-fit point for the normal (inverted) hierarchy is
($\epsilon_{ee}$, $|\epsilon_{e\tau}|$) = (-1.0, 0.0)
((3.0, 1.7))
and the value of $\chi^2$ at this point is
79.0 (78.6) for 50 degrees of freedom, and
goodness of fit is 2.8 (2.7) $\sigma$CL,
respectively.
The best-fit point is different from the
standard case 
($\epsilon_{ee}$, $|\epsilon_{e\tau}|$) = (0, 0),
and this may be because we use only the information
on the energy rate and the sensitivity to NSI
is lost due to the destructive phenomena
between the lower and higher energy bins
(See the discussions in subsect.\,\ref{hk-std}).
The difference of the value of $\chi^2$ for the
standard case and that for the best-fit point
for the normal (inverted) hierarchy
is $\Delta\chi^2=2.7$ (3.4) for 2 degrees of freedom
(1.1 $\sigma$CL (1.3 $\sigma$CL)), respectively,
and the standard case is certainly acceptable
for the both mass hierarchies in our analysis.
From the Fig.\,\ref{fig:fig1} we can read off
the allowed region for
$|\tan\beta|\equiv|\epsilon_{e\tau}|/|1+\epsilon_{ee}|$,
and we conclude that
the allowed region for $|\tan\beta|$
is approximately
\begin{eqnarray}
|\tan\beta|\equiv\frac{|\epsilon_{e\tau}|}{|1+\epsilon_{ee}|}\lesssim 0.8
\qquad\mbox{\rm at 2.5$\sigma$CL}.
\nonumber
\end{eqnarray}

\section{Sensitivity of the Hyperkamiokande atmospheric
neutrino experiment to $\epsilon_{ee}$ and $|\epsilon_{e\tau}|$
\label{atmospheric-hk}}

Let us now discuss the potential sensitivity of HK to
$\epsilon_{ee}$ and $|\epsilon_{e\tau}|$.
Since HK is a future experiment, the simulated
numbers of events are used as ``the experimental data'',
and we vary $\epsilon_{ee}$ and $\epsilon_{e\tau}$
as well as the standard oscillation parameters
trying to fit to ``the experimental data''.
Here we perform an analysis
on the assumption that we know the mass
hierarchy, because
some hint on the mass hierarchy is
expected to be available at some confidence level
by the time HK will
accumulate the atmospheric neutrino data for twenty years.

\subsection{The case with the standard oscillation
scenario
\label{hk-std}}

First of all, let us discuss the case where
``the experimental data'' is the one
obtained with the standard oscillation
scenario.  The values of the
oscillation parameters which are used
to obtain ``the experimental data''
are the following:
$\Delta \bar{m}^2_{31}=2.5\times 10^{-3}\mbox{\rm eV}^2$,
$\sin^2\bar{\theta}_{23}=0.5$,
$\bar{\delta}=0$,
$\sin^22\bar{\theta}_{12}=0.86$,
$\sin^22\bar{\theta}_{13}=0.1$,
$\Delta \bar{m}^2_{21}=7.6\times 10^{-5}\mbox{\rm eV}^2$,
where the bar notation stands for the reference
values for ``the experimental data''.
As in the case of the analysis of the SK data,
we vary the oscillation parameters
$\theta_{23}$, $|\Delta m^2_{32}|$,
$\delta$ and arg($\epsilon_{e\tau}$)
while fixing the other oscillation parameters
$\sin^22\theta_{12}=0.86$,
$\sin^22\theta_{13}=0.1$ and
$\Delta m^2_{21}=7.6\times 10^{-5}$eV$^2$.

\begin{figure}
\vspace*{-5mm}
\includegraphics[scale=0.4]{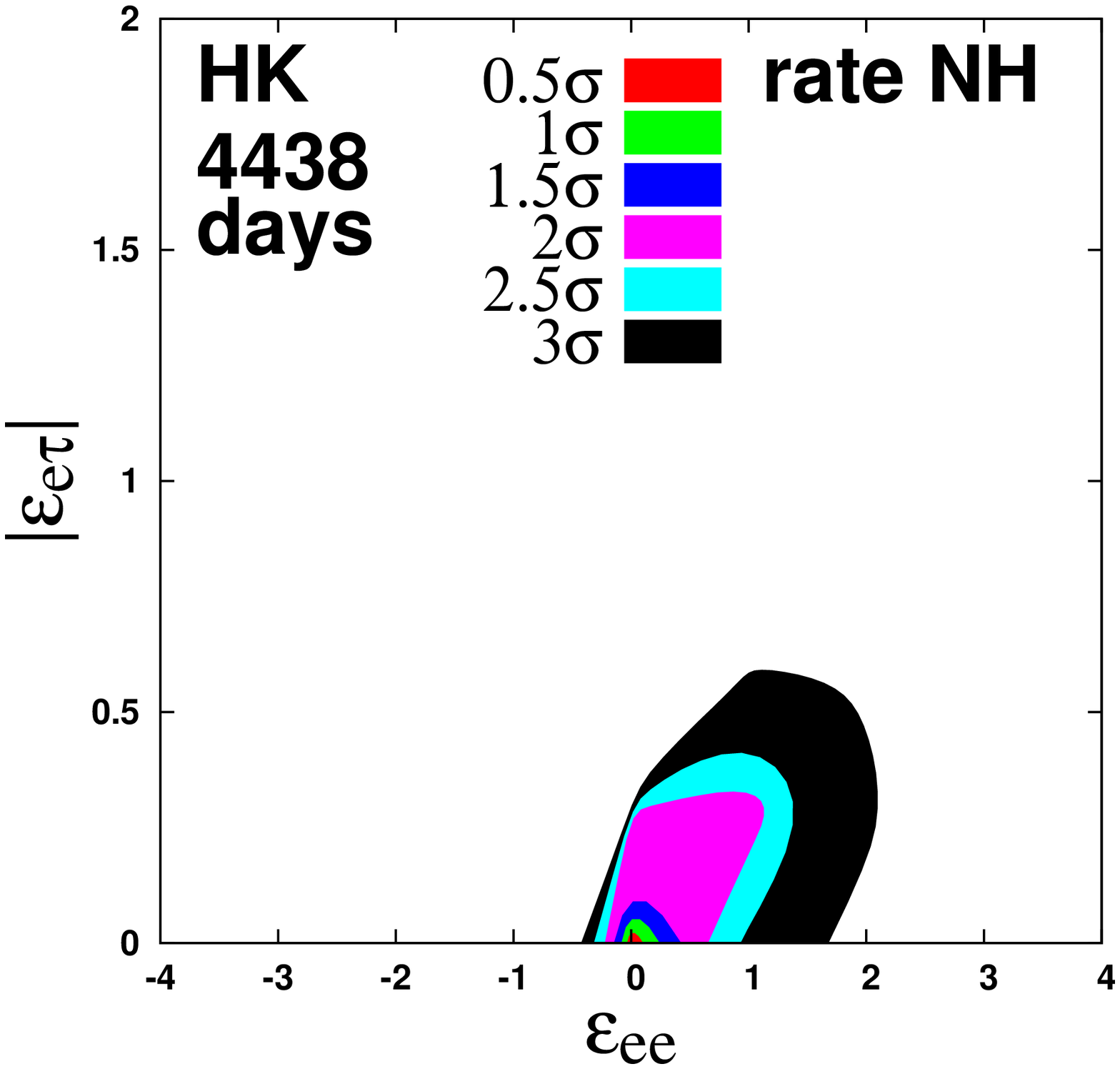}
\hspace*{-19mm}
\includegraphics[scale=0.4]{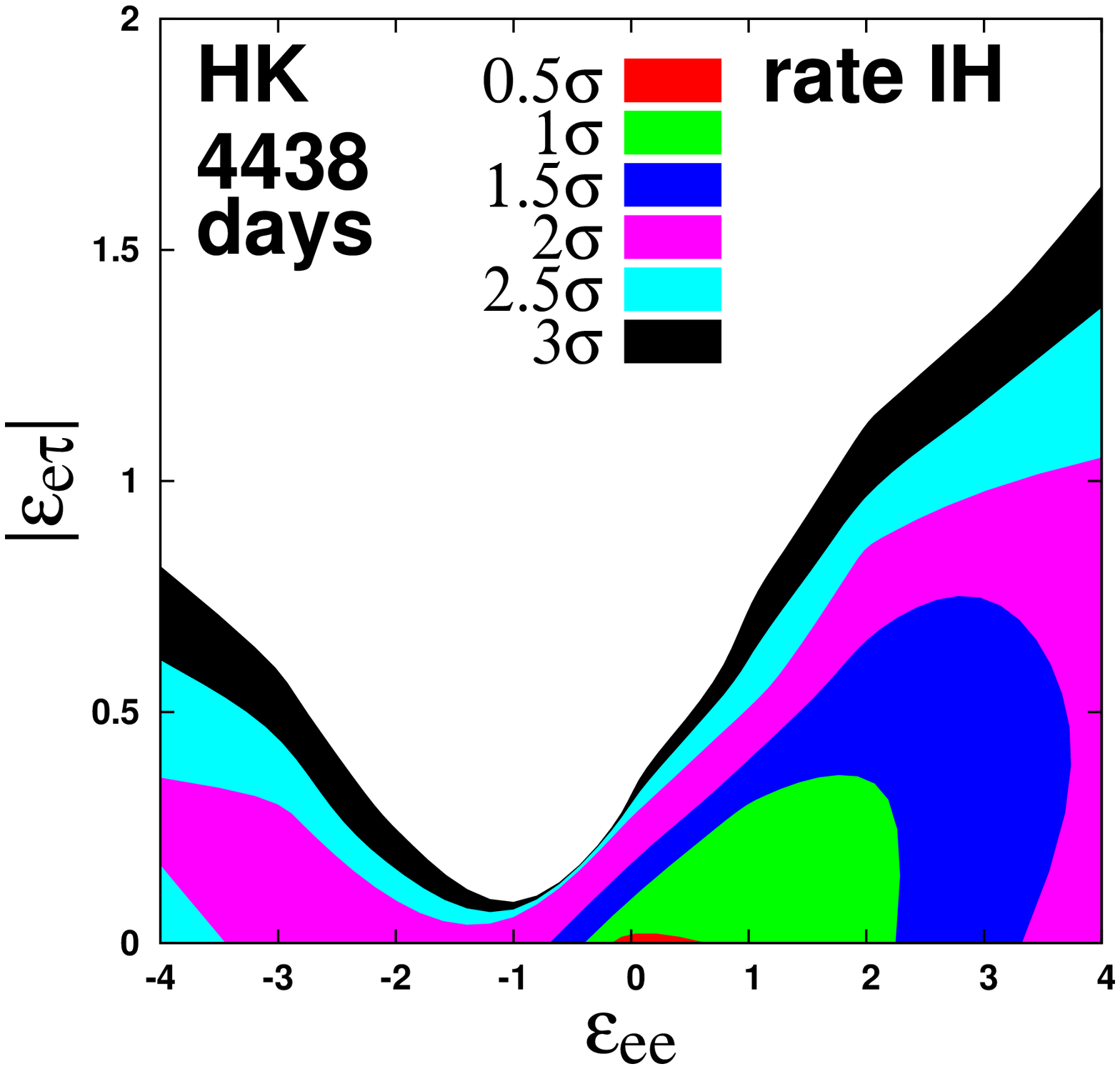}
\includegraphics[scale=0.4]{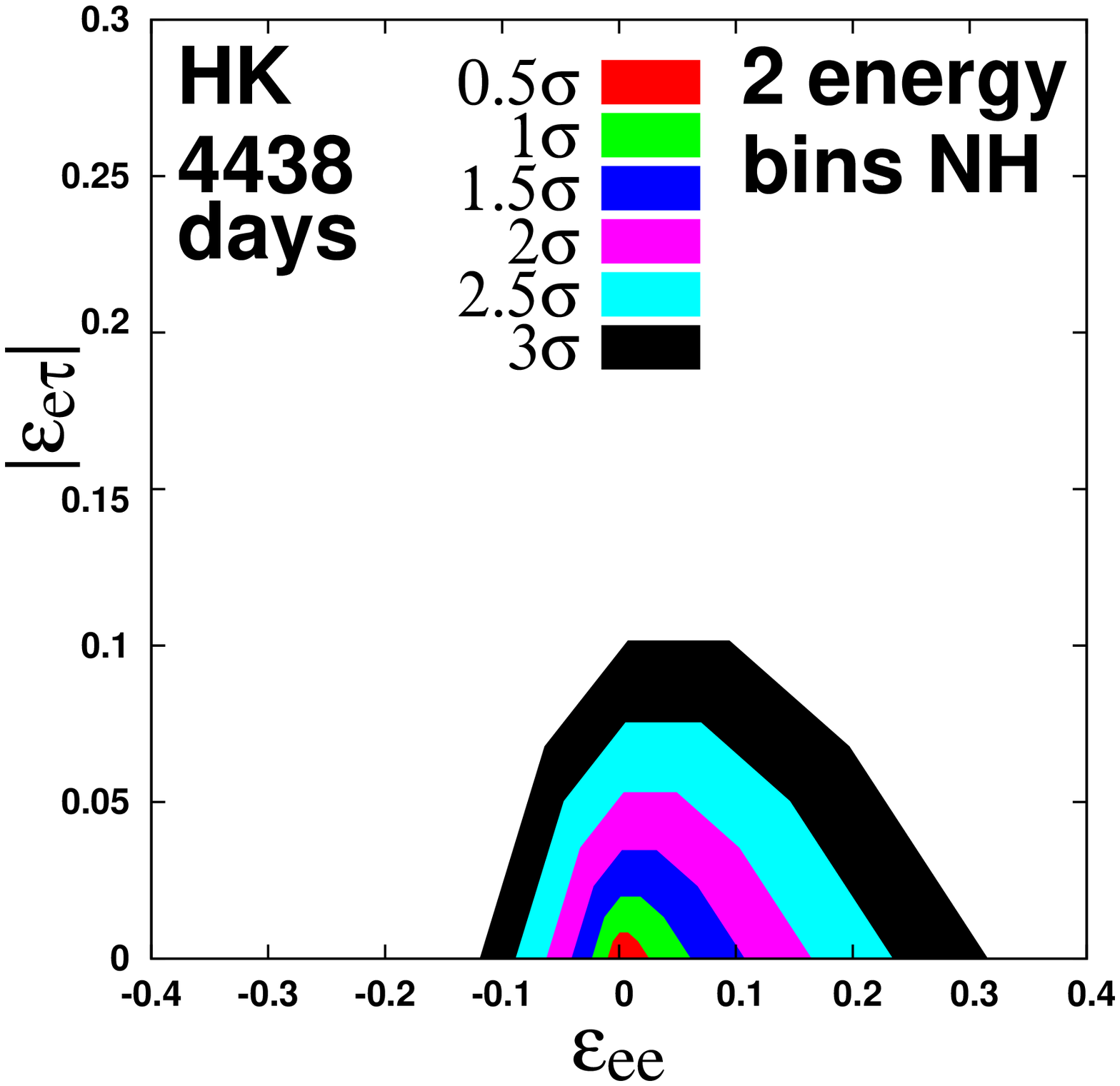}
\includegraphics[scale=0.4]{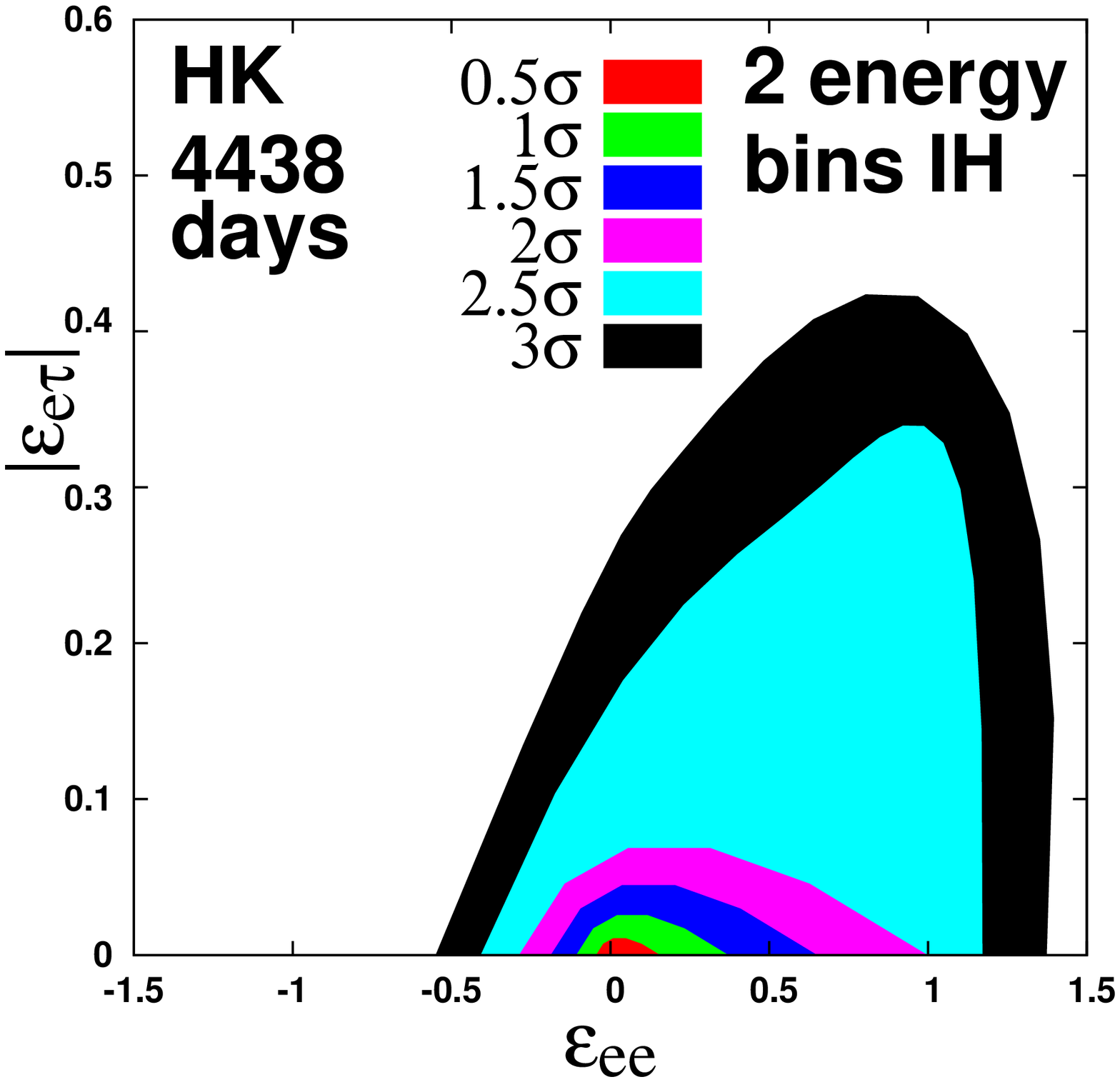}
\vspace*{2mm}
\caption{
The allowed regions
in the ($\epsilon_{ee}$, $|\epsilon_{e\tau}|$) plane
from the HK atmospheric neutrino data for a
normal (inverted) mass hierarchy from the energy-rate
(two energy-bin) analysis.
}
\label{fig:fig2}
\end{figure}

We have performed both
energy rate and energy spectrum (with two energy bins)
analyses,
and the results from the energy rate
(spectrum)
analysis are given by the upper
(lower) panel in Fig.\,\ref{fig:fig2}.
From the energy rate analysis we have
$|\epsilon_{e\tau}/(1+\epsilon_{ee})|\lesssim 0.3$
at 2.5$\sigma$CL.
On the other hand, from the energy spectrum analysis
we get $-0.1\lesssim \epsilon_{ee}\lesssim 0.2$
and $|\epsilon_{e\tau}|< 0.08$
at 2.5$\sigma$ (98.8\%) CL for the normal hierarchy and
to $-0.4 \lesssim \epsilon_{ee}\lesssim 1.2$
and $|\epsilon_{e\tau}|< 0.34$
at 2.5$\sigma$ (98.8\%) CL for the inverted hierarchy.
From Fig.\,\ref{fig:fig2} we note that
the allowed regions
from the energy rate analysis
are much larger than those
from the energy spectrum analysis
for both mass hierarchies,
and that the allowed regions
for the inverted hierarchy
are larger than those
for the normal hierarchy
for both rate and spectrum analyses.
This is because we lose
the significance (particularly
for the high energy bin
in the spectrum analysis and
for the inverted hierarchy case
both in rate and spectrum analyses).
We believe that this is also the reason why
the allowed regions in the SK case
are so large.

\subsection{The case  in the presence of NSI
\label{hk-nsi}}

\begin{figure}
\vspace*{-5mm}
\includegraphics[width=.5\textwidth]{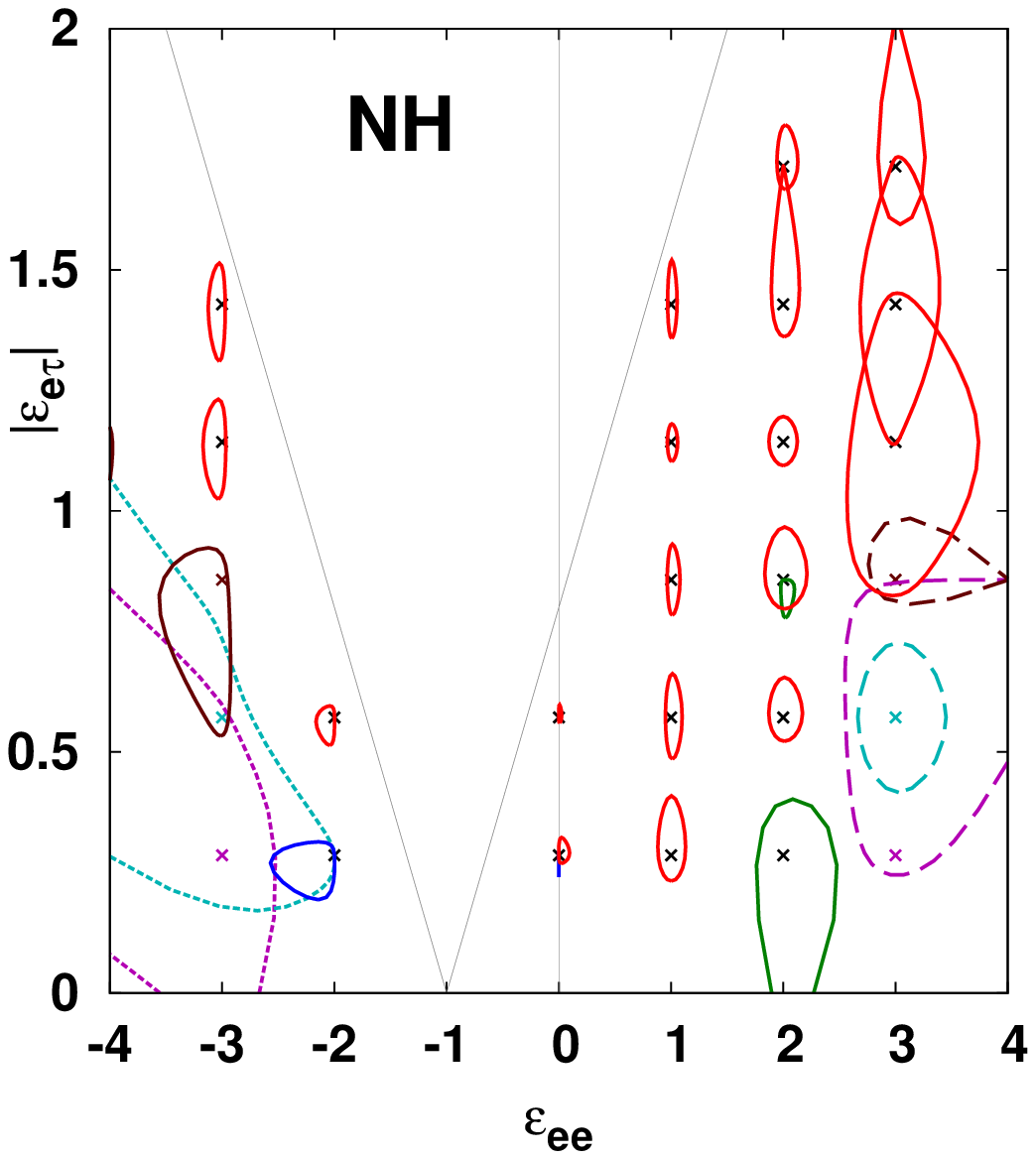}
\hspace*{-8mm}
\includegraphics[width=.57\textwidth]{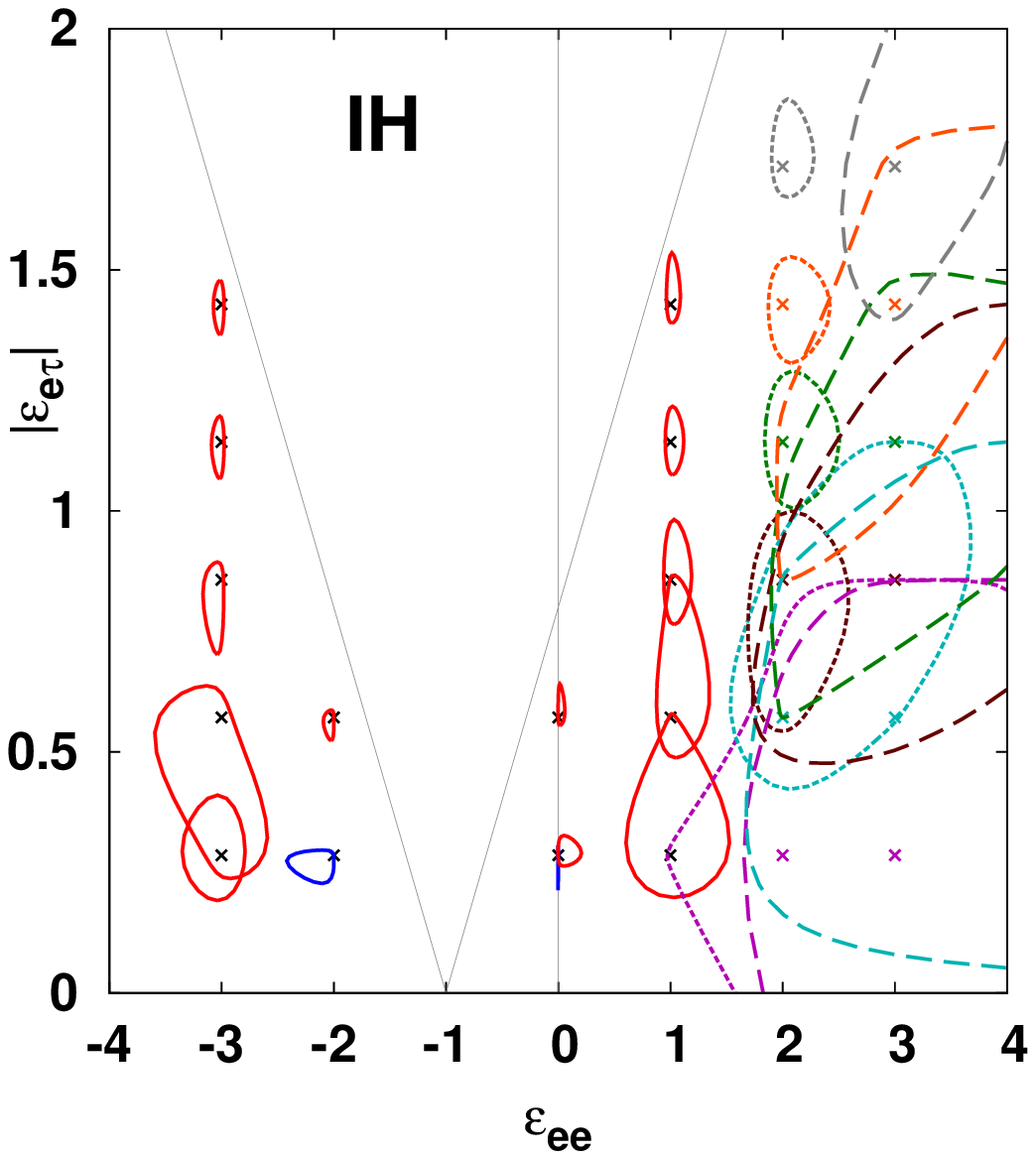}
\vspace*{1mm}
\caption{The allowed region at 2.5$\sigma$CL around the point
$(\epsilon_{ee}, |\epsilon_{e\tau}|) \ne (0,0)$,
where $\bar{\delta}=\mbox{\rm arg}(\bar{\epsilon}_{e\tau})=0$ is assumed.
}
\label{fig:fig4}
\end{figure}

Next let us discuss the case where
``the experimental data'' is the one
obtained with $(\epsilon_{ee}, \epsilon_{e\tau}) \ne (0,0)$.
The analysis is the same as the one in subsect.\,\ref{hk-std},
except that the ``the experimental data'' is produced
assuming the presence of NSI, and here
we perform only an energy spectrum analysis.
The results are shown in Fig.\,\ref{fig:fig4}, where
the allowed regions at 2.5$\sigma$CL
($\Delta\chi^2=8.8$ for 2 degrees of freedom)
around the true points are depicted.
The straight lines $|\epsilon_{e\tau}|=0.8\times|1+\epsilon_{ee}|$
in Fig.\,\ref{fig:fig4} stand for the approximate bound
from the SK atmospheric neutrinos in  Fig.\,\ref{fig:fig1},
and we have examined only the points below
these straight lines.
As seen from Fig.\,\ref{fig:fig4}, 
the errors in $\epsilon_{ee}$ and
$|\epsilon_{e\tau}|$ are small
for $|\epsilon_{ee}| \lesssim 2$
in the case of the normal hierarchy
and for $-3 \lesssim \epsilon_{ee} \lesssim 1$
in the case of the inverted hierarchy.
The errors are large otherwise,
and the reason that the errors are large
is because a sensitivity is lost due to a
destructive phenomenon
between neutrinos and antineutrinos
as was discussed in subsect.\,\ref{hk-std}.

\section{Conclusions
\label{conclusions}}

In this talk we have discussed the constraint
of the SK atmospheric neutrino data on
the non-standard flavor-dependent
interaction in neutrino propagation with the
ansatz (\ref{ansatz}).
From the SK atmospheric neutrino data for 4438 days,
we have obtained the bound
$|\epsilon_{e\tau}|/|1+\epsilon_{ee}|\lesssim 0.8$
at 2.5$\sigma$CL,
while we have little constraint on $\epsilon_{ee}$.

We have also discussed
the sensitivity of the future HK atmospheric neutrino
experiment to NSI by analyses with the energy rate
and with the energy spectrum.
If nature is described by the standard oscillation
scenario, then the HK atmospheric neutrino data
will give us the bound
$|\epsilon_{e\tau}|/|1+\epsilon_{ee}|\lesssim 0.3$
at 2.5$\sigma$CL from the energy rate analysis, and
from the energy spectrum analysis
it will restrict $\epsilon_{ee}$ to
$-0.1\lesssim \epsilon_{ee}\lesssim 0.2$
and $|\epsilon_{e\tau}|< 0.08$
at 2.5$\sigma$ (98.8\%) CL for the normal hierarchy and
to $-0.4 \lesssim \epsilon_{ee}\lesssim 1.2$
and $|\epsilon_{e\tau}|< 0.34$
at 2.5$\sigma$ (98.8\%) CL for the inverted hierarchy.
On the other hand, if nature is described by
NSI with the ansatz (\ref{ansatz}), then
HK will measure the NSI parameters
$\epsilon_{ee}$ and $|\epsilon_{e\tau}|$
relatively well
for $|\epsilon_{ee}| \lesssim 2$
in the case of the normal hierarchy
and for $-3 \lesssim \epsilon_{ee} \lesssim 1$
in the case of the inverted hierarchy.
It is important to use
information on the energy spectrum
to obtain strong constraint,
because a sensitivity to NSI
would be lost due to
a destructive phenomena between
the low and high energy events.

\section*{Acknowledgments}
This research was partly supported by a Grant-in-Aid for Scientific
Research of the Ministry of Education, Science and Culture, under
Grants No. 24540281 and No. 25105009.

\end{document}